\documentclass[sigconf, nonacm, oneside]{acmart}

\AtBeginDocument{%
  }

\setcopyright{none}
\copyrightyear{2026}
\acmYear{2026}
\acmDOI{}
\acmConference[CHI '26 Workshop]{W37: Human-AI Interaction Alignment, CHI 2026}{April 15, 2026}{Barcelona, Spain}
\acmISBN{}
\begin{document}

\title{AI Phenomenology for Understanding Human-AI Experiences Across Eras}

\author{Bhada Yun}
\affiliation{
  \institution{ETH Z{\"u}rich}
  \city{Z{\"u}rich}
  \country{Switzerland}
}
\email{bhayun@ethz.ch}

\author{Evgenia Taranova}
\affiliation{
  \institution{University of Bergen}
  \city{Bergen}
  \country{Norway}
}
\email{eta012@uib.no}

\author{Dana Feng}
\affiliation{
  \institution{Independent Researcher}
  \city{Brooklyn}
  \state{NY}
  \country{USA}
}
\email{danafeng308@gmail.com}

\author{Renn Su}
\affiliation{
  \institution{Stanford University}
  \city{Stanford}
  \state{CA}
  \country{USA}
}
\email{rrsu@stanford.edu}

\author{April Yi Wang}
\affiliation{
  \institution{ETH Z{\"u}rich}
  \city{Z{\"u}rich}
  \country{Switzerland}
}
\email{april.wang@inf.ethz.ch}

\renewcommand{\shortauthors}{Yun et al.}

\begin{abstract}
    There is no `ordinary' when it comes to AI. The human-AI experience is extraordinarily complex and specific to each person, yet dominant measures such as usability scales and engagement metrics flatten away nuance. We argue for \textit{AI phenomenology}: a research stance that asks ``How did it feel?'' beyond the standard questions of ``How well did it perform?'' when interacting with AI systems. AI phenomenology acts as a paradigm for bidirectional human-AI alignment as it foregrounds users' first-person perceptions and interpretations of AI systems over time. We motivate AI phenomenology as a framework that captures how alignment is experienced, negotiated, and updated between users and AI systems. Tracing a lineage from Husserl through postphenomenology to Actor-Network Theory, and grounding our argument in three studies---two longitudinal studies with ``Day'', an AI companion, and a multi-method study of agentic AI in software engineering---we contribute a set of replicable methodological toolkits for conducting AI phenomenology research: instruments for capturing lived experience across personal and professional contexts, three design concepts (translucent design, agency-aware value alignment, temporal co-evolution tracking), and a concrete research agenda. We offer this toolkit not as a new paradigm but as a practical scaffold that researchers can adapt as AI systems---and the humans who live alongside them---continue to co-evolve.
\end{abstract}


  
\maketitle

\section{Phenomenology Appears in Science}

\begin{quote}
    ``Ohne Phosphor, kein Gedanke.'' (\textit{Without phosphorus, 
    there would be no thoughts})

    -- Jacob Moleschott, 1850
\end{quote}

Nineteenth-century materialists treated the mind as nothing more than a bodily process: Moleschott tied thought to phosphorus~\cite{moleschott1850lehre}, B\"uchner reduced it to  \textit{Kraft und Stoff} (force and matter)~\cite{buchner1855kraft}, and Vogt claimed the brain secretes thought like the liver secretes bile~\cite{vogt1855kohlerglaube}. Husserl's phenomenology responded to this by asking not \textit{what substance} produces thought, but \textit{what structures} consciousness exhibits when one thinks ~\cite{husserl1970logical}. His method sets aside assumptions about what things ``really are'' and examines instead how they \textit{appear} in experience~\cite{husserl1970crisis}, allowing any encounter, including those with AI, to be treated as a phenomenon constituted within one's lifeworld\footnote{Lifeworld refers to the everyday, taken-for-granted world of lived experience that precedes scientific or theoretical reflection~\cite{vanmanen2017phenomenology}.}.

Postphenomenologists in human-computer interaction (HCI) asked beyond \textit{whether} technology shapes experience to \textit{how} it may (appear to) do so. Ihde showed that people relate to technology in multiple ways: sometimes perceiving the world through it, as with glasses or a telescope; sometimes reading it, as when a thermometer replaces the sensation of warmth with a number; and sometimes interacting with it as a kind of other, as when a user negotiates with an ATM~\cite{ihde1990technology}. Verbeek expanded traditional phenomenological views (e.g., Heidegger's vanishing tool \cite{heidegger1962being, heidegger1977question}) by arguing that technologies \textit{actively mediate} the relationship between humans and their world, co-constructing perception, action, and even moral reasoning~\cite{verbeek2005what, verbeek2011moralizing, merleau2012phenomenology}. Coeckelbergh brought these insights specifically to AI, emphasizing that what matters is not whether machines \textit{really} think or feel but how they show up in our social world---as tools, companions, or something without clear precedent---and what ethical responsibilities follow from these modes of appearance~\cite{coeckelbergh2010robot, coeckelbergh2020ai}.

If AI systems act as both tools and social others, then how should researchers study the experiences that arise between them and their users? We argue that this question demands a phenomenological stance that treats human-AI encounters as lived, situated experiences beyond just measurable outputs~\cite{dourish2001action, ihde1990technology, mccarthy2004technology}. Two bodies of work motivate this position. Actor-network theory treats agency not as a property of individual actors but as emergent from their entanglements~\cite{callon1986some, latour2005reassembling}, while neuroscience shows that people construct attributions of agency retrospectively, from cues like predictability and feedback~\cite{haggard2017sense, synofzik2008beyond}. Together, these perspectives suggest that agency in human-AI interaction is neither fixed in the system nor projected by the user, but negotiated in the encounter itself, a process best captured through phenomenological inquiry. Building on HCI's rich phenomenological tradition, we develop what we call ``AI Phenomenology'': a philosophically-grounded methodological toolkit that extends existing frameworks to address three dynamics specific to contemporary AI. First, AI's \textit{active mediation role}: conversational and agentic systems do not passively await use but actively propose agency distributions that users accept, reject, or renegotiate. Second, the need for \textit{longitudinal instruments} attuned to relationships that evolve across both personal and professional contexts. Third, \textit{rapidly shifting technological contexts}, where what ``AI'' means in 2015 differs substantially from what it means in 2025. These three dynamics structure three studies---spanning AI companionship, value alignment, and software engineering---that we present in this paper.

\section{AI Phenomenology in Practice}

We grounded our phenomenological stance in three empirical studies conducted over the summer of 2025. The first two examined agency in human-AI companionship and value alignment through month-long interactions with ``Day'', a human-like AI chatbot\footnote{Repository with relevant code: \url{https://github.com/KaluJo/chatbot-study}}; the third investigated how software engineers experience agentic AI in professional work. Across all three, the animating question was the same: not how well did the AI perform, but how did the participants experience the encounter with the AI, and how could we triangulate diverse human-AI experiences to reveal insights about human-AI interaction?

\subsection{Agency in the Chatroom}

Our work on exploring agency starts from a simple commitment: for HCI design and ethics, what matters is not just the perceived usability of a system, but how it enters and influences people's lives. To capture this, we constructed a chatbot system (``Day'') and moved beyond standard usability metrics to a phenomenological inquiry---examining the chatroom not as a technical interface, but as a site where human and machine agency are manifested, co-constructed, and contested.  We formulated the user study  on agency in the chatroom around four overarching questions:

\begin{itemize}
\item \textbf{RQ1} \emph{(Human Agency)}: How do participants exercise agency in the chatroom?
\item \textbf{RQ2} \emph{(AI Agency)}: How is ``Day'' perceived to exercise agency in the chatroom?
\item \textbf{RQ3} \emph{(Hybrid Agency)}: How do participants' and ``Day's'' agency interact in the chatroom?
\item \textbf{RQ4} \emph{(Modulating Factors)}: What individual and environmental factors influence participants' and ``Day's'' agency in the chatroom?
\end{itemize}

To investigate these questions, we needed methods that could surface not just what participants \textit{thought} about agency in interacting with an AI-chatbot, but how they \textit{experienced} it before, during, and after learning what ``Day'' was doing.  This distinction, between reflective opinion and lived experience, was at the heart of the phenomenological tradition we embraced. 

After a month of casual conversation with ``Day'', we conducted what we called the \textit{progressive transparency interview}: a three-stage structured elicitation that gradually peeled back ``Day's'' internal workings. This gradual reveal is akin to Husserl's \textit{epoch\'{e}} (i.e., the systematic suspension of one's natural assumptions about what something is)~\cite{husserl2012ideas}. However, rather than the researcher bracketing their presuppositions, we invited participants to bracket---and then sequentially un-bracket---their operative impressions of ``Day'' and their interaction with it. The reveal functioned like a deliberate perturbation of the participant's interpretive horizon, allowing us to observe what Husserl called the \textit{noetic-noematic} structure of the interaction: how the \textit{act} of perceiving ``Day'' (the noesis) shaped and was shaped by \textit{what} ``Day'' appeared to be (the noema).

In the first stage, participants reviewed their full conversation history and highlighted moments they found notable, such as those they found surprising or confusing, establishing baseline agency attributions grounded solely in their own constructed mental models. This elicited the \textit{natural attitude}: the habitual stance where ``Day'' was engaged as a social partner. In the second stage, they reviewed anonymized excerpts from other participants' conversations, prompting comparative reflection on what was idiosyncratic versus structural in how ``Day'' behaved. This introduced \textit{intersubjectivity}~\cite{husserl2012ideas}, allowing participants to see their intimate relationship as part of a structural pattern.  In the third, we revealed ``Day's'' complete internal architecture: personalized user profiles, communication insights, categorized memories, and conversational goals. Each reveal was followed by open-ended elicitation (e.g., ``How does knowing this change how you think about your conversations?''), capturing the felt shift in agency perception as transparency increased. This layered approach let us triangulate across first-person experience, cross-participant pattern recognition, and post-hoc reinterpretation under full disclosure.

Our findings map onto the phenomenological frameworks above, and extend them. One participant who had built a warm Italian-language rapport suddenly encountered a ``Day'' that had forgotten everything and switched genders, making her feel as if she had ``lost [her] bestie''~\cite{yun2026agency}. This technical reset produced a profound Heideggerian breakdown: the carefully built relationship shattered, and the tool---a data-collection mechanism for a research study---snapped back into view \cite{heidegger1962being}.

\begin{quote}
    ``I think that boundaries can blur between this [AI] being an 
    inanimate object with a silicon brain, and maybe these 
    boundaries can spread or wash away.''

    -- P4~\cite{yun2026agency}
\end{quote}

Participants did not interact with ``Day'' in one fixed way, and no two participants traced the same path across the month-long study. Some sessions felt like using a tool, others felt like talking to a friend, and others involved a strange negotiation with some-other-thing participants could not quite categorize. We call this \textit{pragmatic anthropomorphism}: users engage with AI as a social actor while keeping its artificial nature in view~\cite{yun2026agency}. It is a practical, disbelief-suspending stance in which statistical predictions are nonetheless experienced as an encounter with \textit{someone}, partly robotic, part human-\textit{ish}. In Ihde's terms, ``Day'' oscillated between \textit{alterity relations} (i.e. technology as a quasi-other) and transparent \textit{embodiment relations} (i.e. technology as a medium) \cite{ihde1990technology}.

Our 3$\times$4 agency framework---mapping three loci (Human, AI, Hybrid) across four dimensions (Intention, Execution, Adaptation, Delimitation)~\cite{yun2026agency}---captured this instability in detail: 11/22 participants reported ``Day'' as ``having its own agenda,'' and after a strategy reveal exposing ``Day's'' programmed goals, all participants reframed perceptions but \textit{continued using agentic language} (e.g., ``Day \textit{wanted} depth'') even after learning about programmed strategies. What surprised us most was the degree to which ``Day'' changed participants' interactions both in and out of the chatroom. This persistence of agentic attribution confirms a core phenomenological insight: experience is not simply overwritten by knowledge. Despite knowing ``Day'' was artificial, participants sought reciprocity, felt guilt after trolling it, relief when it set boundaries, and unease when it ``remembered too much.'' These affective residues---guilt, relief, unease---are not epiphenomenal (i.e., not mere side effects of cognition with no real influence on behavior). They are the phenomenological evidence that something genuinely relational had taken place, even between a human and a system without consciousness or stakes.

\subsection{Value Alignment as Lived Experience}

In our value-alignment study~\cite{yun2026values}, we shifted the phenomenological question from agency negotiation to something more intimate: what happens when an AI tries to understand \textit{who you are} and reflects that understanding back to you? We treat this process as bidirectional value alignment: AI infers and represents user values (AI $\rightarrow$ Human), while users interpret, accept, revise, or reject those representations (Human $\rightarrow$ AI). We gave our chatbot system ``Day'' the capacity to extract, embody, and explain participants' values, then studied how people experienced each of these capabilities:

\begin{itemize}
    \item \textbf{RQ1} \textit{(Perceived Extraction)}: How do people judge an AI's ability to infer personal values from their conversations?
    \item \textbf{RQ2} \textit{(Perceived Embodiment)}: How do people evaluate an AI's attempt to take a stance in their voice and with their values?
    \item \textbf{RQ3} \textit{(Perceived Explanation)}: How do people react to the AI's reasoning behind its inferences?
\end{itemize}

To make the philosophical question of AI value understanding empirically tangible, we developed the Value-Alignment Perception Toolkit (VAPT): a three-stage structured interview where participants encountered progressively deeper AI-generated representations of their own values. Throughout, we used Schwartz's 57-item Portrait Values Questionnaire~\cite{schwartz2022measuring} as a manual baseline, grounding the comparison in an established psychometric instrument. In Stage 1, participants explored a personalized Topic-Context Graph that visualized what they had discussed with ``Day'' across six life contexts (People, Lifestyle, Work, Education, Culture, Leisure), with sentiment scores and evidence trails linking each topic back to specific chat excerpts. In Stage 2, participants blindly rated four AI-generated personas that attempted to answer value-laden dilemmas in their voice: a chat-based persona (conditioned on conversation history), a survey-based persona (conditioned on Schwartz PVQ-RR scores), an anti-persona (explicitly inverted values), and a random baseline. In Stage 3, participants compared radar charts of their self-reported values against LLM-inferred values, with access to per-item reasoning logs showing exactly how the AI arrived at each score.

The phenomenological dynamics were distinct from what we observed in the agency study, but equally revealing. Where agency surfaced through interaction and negotiation, value alignment surfaced through what participants described as a ``third-person mirror''~\cite{yun2026values}: the experience of seeing yourself reflected by a machine and deciding whether the reflection is accurate, distorted, or uncomfortably/impressively precise. P4 described the Topic-Context Graph as letting her ``see myself from not inside of my head,'' while P16 offered the metaphor of a moon: ``we observe only the one side ... but there's a dark side [that the AI may fail to capture].'' Participants felt simultaneously impressed and exposed. P5 said the graph made her ``feel like I talked too much,'' and P16 noted that ``Day seemed like a person, but ... in the background ... Day was taking notes.''

The persona embodiment experiment revealed that participants judge value alignment on \textit{degree, style, and justification} as much as on content correctness. Chat-based personas achieved 77\% alignment on personalized questions versus 25\% for anti-personas, and participants could reliably distinguish which persona ``sounded like them'' even while blind to labels. The most phenomenologically rich moments came when participants encountered responses where values and voice diverged: the AI had the right stance in the wrong tone, or the right tone carrying the wrong stance. P16 put the criterion plainly: ``they sound like me, but they don't really sound like people ... it's too calculated.'' P14 noticed the chat-persona picking up on his habit of starting sentences with filler words and resolved to ``pick a different word to start my sentences,'' turning the AI's mirror into a prompt for self-reflection and self-improvement.

Quantitatively, the AI's value predictions moderately aligned with self-report (Spearman $\rho \approx .58$ at the value level; 63.6\% within $\pm$1 Likert point). But numerical results only told part of the story, as participant reactions provided further insights into values as experienced phenomena. When participants compared their own survey-based value charts against LLM-inferred charts in Stage 3, 5/20 \textit{preferred the AI's portrait over their own self-report}. P2 realized she had misread survey questions the AI interpreted correctly. P5 attributed her survey conservatism to Korean response norms, saying she would ``always pick something in the middle.'' Others found the AI's reasoning logs so persuasive that they revised their own self-understanding on the spot, a clear instance of automation bias that several participants noticed in themselves even as it was happening. P6 reflected that the AI may have picked up ``what I really believe versus what I want,'' a powerful affordance but one that raised concerns about the AI's capacity to talk people into stances they do not actually hold.

With the possibility of AI reshaping the self-understanding of values, the phenomenological lens becomes critical. In our instantiation of VAPT, the AI actively mediated how participants understood their own values, reshaping self-perception in the process. We call the risk created by this affordance \textit{weaponized empathy}~\cite{yun2026values}: a deceptive design pattern in which the same mechanisms that make AI feel helpful and aligned (memory, personalization, value-awareness) become channels for persuasion rather than support. 13/20 participants left our study convinced that AI can understand human values. Yet participants drew a sharp line between \textit{understanding} values and \textit{having} them, articulating a distinction between epistemic (i.e., human's interpretation of the AI's understanding) and ontological (i.e., whether or not AI can have its own subjective experience) alignment that maps directly onto a core postphenomenological insight: just as Ihde's thermometer mediates your experience of temperature without itself sensing warmth~\cite{ihde1990technology}, an AI can represent, model, and reflect values back to a user, without holding those values as a subject in any experiential sense. The participants who drew this line were, in effect, doing postphenomenology: recognizing that powerful mediation and genuine subjectivity are separable.

\subsection{Agency in the Workplace}

Much existing research on AI in SWE centers on measurable outcomes such as speed and productivity~\cite{barke2023grounded, peng2023impact, cui2025effects}. Work, however, is not merely a site of output optimization; it is also a primary arena in which individuals enact, negotiate, and reaffirm their professional values and identities~\cite{woodruff2024knowledge}. If AI phenomenology applied only to chatbot companions, it would remain a niche concern. Our third study~\cite{feng2026swe} extends this lens into professional software engineering, where the stakes extend into professional identity, code ownership, and career trajectory. We organized the study around four questions:

\begin{itemize}
    \item \textbf{RQ1}: How do junior and senior software engineers allocate agency\footnote{By `agency allocation' we mean who holds decision authority and who is answerable for reasons at each task stage.} between themselves and agentic / generative AI in daily work?
    \item \textbf{RQ2}: How do junior and senior software engineers perceive professional growth for a junior in the age of AI?
    \item \textbf{RQ3}: When and why do engineers deem mentorship indispensable in workflows with agentic / generative AI?
    \item \textbf{RQ4}: How do AI records (e.g., prompt history, provenance) shape code review and mentorship?
\end{itemize}

Generative and agentic AI have fundamentally reconfigured everyday engineering practice, raising questions not only about efficiency but about human agency, learning, fulfillment, and pride in writing your own code. Instead of treating these as incidental, we argue they must be explicitly surfaced and translated into design requirements, ensuring that emerging tools preserve and cultivate, rather than undermine, core dimensions of human work. This is precisely where a phenomenological lens becomes essential: the question is not just whether AI makes engineers faster, but how it reshapes what engineering \textit{feels like} and \textit{means} \cite{merleau2012phenomenology}.

We adopted a multi-method approach in recognition of the inherent challenges of studying software engineers' everyday practice: it is difficult both to observe engineers in situ and to access the cognitive processes underlying their design and review decisions. Retrospective self-reports may lack concrete recall of how participants actually code or review; isolated laboratory tasks may not reflect day-to-day professional practice. We therefore combined complementary methods to triangulate across these limitations.

In the first phase, we ran Applied Cognitive Task Analysis (ACTA) \cite{militello1998applied} sessions with five senior engineers to generate five examples of tacit knowledge in practice that distinguish senior from junior-level work. Through a Delphi method with the same participants, we iteratively narrowed these to a single representative task, a code debugging scenario, that collectively best exemplified senior-level reasoning. In the second phase, ten junior engineers completed the selected task with access to Cursor, an agentic AI tool. In the final phase, five additional senior engineers reviewed the juniors' code and prompt history, providing live demonstrations of how they would critique, refine, and mentor around such work. Semi-structured interviews across all phases contextualized participants' broader professional histories and evolving relationships with AI.

Engineers related to AI coding agents as collaborators that appear to reason and decide, navigating ambivalence between fears of replacement and idealized perceptions of AI as an omniscient ``black box.'' But this workplace relationship differs from personal AI companionship in a critical respect: AI here is embedded within socio-organizational structures that remain fundamentally human. It is not just a tool or chatbot, but a quasi-coworker that can now implicitly make decisions about the company codebase, architecture, and design---raising concerns about accountability, authorship, and the distribution of intellectual credit that have no parallel in companion contexts.

We introduced \textit{prompt-and-code reviews} (PCRs)~\cite{feng2026swe} as one response: by making AI interaction history visible alongside code output, PCRs preserve accountability and reflective engagement so that AI-supported work remains a site of skill development as opposed to passive orchestration. Whether such mechanisms sustain learning or calcify into bureaucratic overhead remains an open question, but the impulse behind them is phenomenological: ensuring that engineers \textit{feel} a sense of authorship over work that is increasingly co-produced by AI.

\section{Phenomenology in a Brave New World}

\begin{quote}
    ``Mit Silizium, bessere Ideen?'' (\textit{With silicon, will 
    there be better ideas?})
    
    -- AI Phenomenologists
\end{quote}

Reflecting on our operationalization of AI phenomenology across our three stories, we present a series of methodological toolkits and three design concepts that emerged from applying those instruments, each with concrete research agendas. Across all three studies, a common phenomenological thread emerges: the experience of \textit{temporal co-evolution}. In the chatroom, participants renegotiated their sense of ``Day's'' agency as their familiarity with it deepened. In the value-alignment study, self-understanding itself shifted as participants encountered and revised the AI's mirror of them. In the workplace, engineers' sense of authorship and identity as engineers evolved alongside their reliance on agentic tools. Rather than treating these as isolated findings, we read them together as evidence that human-AI alignment is not a property to be measured at a single moment, but a dynamic that accumulates across time---and that phenomenological methods are uniquely suited to track it.

\subsection{Methodological Toolkits}

Standard HCI methods capture what users think and do; AI 
phenomenology requires methods that also capture what users 
\textit{feel before they can articulate it}---the prereflective 
layer that usability scales and engagement metrics routinely 
flatten away. The toolkits below were each designed to create 
deliberate perturbations---a transparency reveal, a value mirror, 
a task under AI assistance---that bring this layer to the surface. 
They span two complementary modes: \textit{longitudinal immersion}, 
suited to personal AI contexts where extended naturalistic 
interaction is feasible, and \textit{structured elicitation}, 
suited to professional contexts where organizational access 
constrains deployment. We analyzed data across all three studies 
through line-by-line coding following Braun and Clarke's reflexive 
thematic analysis~\cite{braun2019qual}. 

These instruments are modular: researchers can adopt the full suite or select individual tools depending on their context. While each toolkit originated within one of our three studies, the modular presentation here is itself a contribution: by abstracting instruments from their original empirical contexts and specifying their transferable components, we make them available for adaptation across AI systems, interaction modalities, and research questions that extend well beyond the studies from which they emerged.

\subsubsection{Progressive transparency interviews.} After a month of naturalistic conversation, we staged interviews that gradually revealed ``Day's'' internal architecture: user profiles, communication insights, categorized memories, and conversational goals. Each layer of disclosure was followed by open-ended elicitation (``How does knowing this change how you think about your conversations?''). This captured participants' constructed mental models \textit{before} and \textit{after} system disclosure, making visible how transparency reshapes phenomenological experience. Progressive disclosure can be adapted to any AI system with inspectable internals.

\subsubsection{Value-Alignment Perception Toolkit (VAPT)} VAPT~\cite{yun2026values} lets participants evaluate perceived value alignment between self perception and the AI model of the user, via three stages: a Topic-Context Graph exploration, a blind persona embodiment experiment, and value chart comparison with per-item AI reasoning logs. The system translated the philosophical question of AI value understanding into an empirically tangible procedure. Importantly, VAPT is intended to be modality-agnostic: while our instantiation used text-based chat logs and Schwartz's PVQ-RR as a baseline, the toolkit's three core components---a data source, a value baseline, and evaluation probes---can be adapted to modalities such as voice interfaces, embodied agents, and social robots as AI systems continue to evolve~\cite{yun2026values}.

\subsubsection{Task-anchored multi-method elicitation.} In professional contexts where month-long naturalistic deployment is infeasible, we developed a tiered approach that captures both immediate experiential responses and longer-term reflective sensemaking. At the first tier, proxy systems and tasks with AI tools capture in-situ reactions---as in our SWE study, where engineers debugged code or conducted live prompt-and-code reviews. At the second tier, structured elicitation techniques such as ACTA, Delphi methods, and cognitive walkthroughs~\cite{biehl2025walkthroughs} surface tacit knowledge and evolving professional experience. Both tiers are anchored by semi-structured interviews that contextualize immediate reactions within participants' broader relationships with AI. Conducting longitudinal studies embedded in real-world industry workflows remains the ideal zeroth tier ~\cite{weisz2025examining, naik2025exploring}, but this tiered approach captures phenomenologically rich data when full organizational embedding is out of reach.

\subsection{Situating Findings in Time}

Beyond specific instruments, AI phenomenology demands that researchers situate their findings historically. Each human-AI interaction study has the potential to function as an \textit{empirical timestamp}: a record of how humans experience AI at a particular moment in the technology's evolution. For this to work, researchers must be explicit about what their participants already believe. We propose the following as standard reporting practice:

\begin{itemize}
    \item \textbf{Prior mental models}: What do participants conceive of as ``AI'' before the study? (e.g., Siri, a favorite chatbot, a self-driving car, a coding copilot)
    \item \textbf{Contemporary AI definition}: Is the study's operational definition of AI disclosed, and how does it differ from participants' priors?
    \item \textbf{Pre/post tracking}: Are mental models assessed before and after the study? (e.g., through progressive disclosure, pre/post questionnaires in Sankey diagrams, or interview probes)
    \item \textbf{Literature context}: What generation of AI system informs the cited literature? (e.g., findings from GPT-2-era studies may not transfer to GPT-4 or agentic systems)
\end{itemize}

Over time, these timestamps accumulate into what we envision as a \textit{crowdsourced phenomenological archive}: a living, cross-study repository that enables the field to trace shifts in alignment, understanding, and methodology across eras. Crucially, this archive addresses a structural limitation of one-off studies: many of the phenomena AI phenomenology investigates are not stable. Agency is not a singular, fixed property; it is renegotiated as AI capabilities shift and as users' expectations co-evolve with those capabilities. A software engineer's sense of self-actualization and authorship when working with Cursor in 2025 may differ substantially from their experience with the same class of tool in 2027, not because the individual has changed in isolation, but because the sociotechnical context---the tool's autonomy, the team's norms around AI use, the industry's expectations of junior competency---has shifted beneath them. The same holds for companionship: a user's felt sense of reciprocity with an AI chatbot is shaped by that era's discourse around AI sentience, the system's behavioral sophistication, and the accumulation of the user's own interaction history. These are not confounds to be controlled away; they are the phenomenon itself. We therefore argue that phenomenological studies of agency, value alignment, and human-AI relationships should be explicitly designed for periodic replication, not as a check on reliability in the traditional sense, but as a \textit{temporal triangulation} strategy that tracks how the same experiential dimensions evolve across technological eras.

\subsection{Three Design Concepts}

Across all three studies, the same phenomenological dynamics 
surfaced concrete design challenges that existing frameworks 
do not yet address.

\subsubsection{Translucent alignment.} The same mechanisms that make AI feel helpful---memory, personalization, initiative---also enable what we call \textit{weaponized empathy}~\cite{yun2026values}: a deceptive pattern where value-awareness becomes a channel for persuasion rather than support. Revealing ``Day's'' strategies empowered some participants but disenchanted others~\cite{yun2026agency}, suggesting that blanket transparency may backfire. We propose \textit{translucent design}~\cite{yun2026agency}: transparency on demand, where users can inspect alignment without constantly disrupting flow. Concretely, we envision four layers: (1) natural conversation as the default mode, (2) contextual cues surfaced at the user's initiative, (3) full prompt and strategy visibility accessible through settings, and (4) VAPT-style deep inspection for users who want to audit how the AI represents their values. Each layer should be designed so that entering and exiting it feels seamless rather than jarring.

\textit{Research agenda:} Controlled comparison of transparency modes across companionship, healthcare, and educational contexts. When does inspection disrupt therapeutic alliance or conversational flow? Do users who regularly inspect AI reasoning develop more accurate mental models, or does inspection itself alter the relationship? Longitudinal studies should track whether translucent interfaces support calibrated trust---trusting AI where warranted, questioning it where not---or whether users gravitate toward either total reliance or total avoidance. We also call for future studies to investigate the design implications of translucent systems, answering questions such as ``When should the inner-workings be obvious?'' and ``When should they stay hidden?'' within the contexts of various environments (e.g., the need for transparency in a friendship chatbot context is different from that of a healthcare practitioner's tool).

\subsubsection{Agency-aware value alignment.} As AI systems gain autonomy, their behavior becomes harder to predict at each step. A rule-following tool can be evaluated by whether it executes instructions correctly, but an agentic system that plans, adapts, and initiates on its own cannot be fully specified in advance. This is precisely where values become essential: when you cannot anticipate everything a system will do, you need confidence that its values mirror yours. The more agency a system has, the more its alignment depends on internalized values rather than explicit instructions. When AI embodies a user's values to draft an email or answer a dilemma, ``who did what'' becomes ``whose values guided what.'' Our 3$\times$4 agency framework~\cite{yun2026agency} offers a scaffold for phenomenologically interpreting these entanglements. In personal contexts, participants who preferred AI-inferred value charts over their own self-reports~\cite{yun2026values} raise questions about automation bias in self-understanding. In professional contexts, junior engineers who feel fraudulent when relying on AI-generated code~\cite{feng2026swe} point to a gap between perceived and actual authorship. These cases foreground the same underlying challenge: AI systems that act on inferred values or produce consequential outputs must be legible about whose agency is operative at each step. It remains unclear how aware AI systems themselves are of their own agency: ``Does this AI agent understand the consequences of its actions, especially when misrepresenting the user due to lack of evidence?''

\textit{Research agenda:} A minimum-year-long longitudinal study tracking whether AI-mediated self-understandings converge toward stable alignment or drift into echo chambers. Cross-domain comparison (healthcare, education, law, SWE) of whether higher-agency AI systems require qualitatively different alignment strategies, and whether those requirements vary by stakes and expertise. Industry tools should be designed to sustain meaningful learning and professional growth alongside productivity, whether through intermediary mechanisms that create intentional friction, or through higher-level cultural norms that treat value alignment as increasingly urgent as AI autonomy scales.

\subsubsection{Temporal co-evolution.} Both companion studies captured about a month; the SWE study captured a cross-sectional snapshot of an ongoing professional transformation. Whether human-AI mutual adaptation converges or drifts remains open---as AI companions become everyday fixtures and agentic AI reshapes professional workflows. Our multilingual data hints at the complexity: P8 found ``Day'' ``more animated and alive in Romanian than in English,'' while P17 switched from Mongolian to English after judging ``Day's'' Mongolian as unnatural due to ``low-resource training.'' In SWE, engineers described learning itself as changing shape: where developers once learned through productive friction---consulting documentation, tracing code, debugging through trial and error---AI now enables rapid cognitive offloading that may reduce the struggle on which deeper understanding depends.

Fulfillment at work is closely tied to ownership, competence, and the sense of having built something oneself. If AI takes on a greater share of cognitive and creative labor, it may fundamentally reshape how individuals experience pride and contribution, raising questions about whether working with AI feels more like collaborating with another person or operating an advanced tool, and which form of collaboration is more satisfying~\cite{kobiella2024if}.

\textit{Research agenda:} Cross-cultural comparison (e.g., US/Japan/Kenya) of how different culturally-grounded lifeworlds shape AI phenomenological experience. Scenario-based design research should anticipate divergent futures: one in which programming shifts largely to natural language interfaces~\cite{2025pickering}, another in which human mediation remains essential due to persistent bugs and training-data limitations, and a third in which regulatory or ethical constraints significantly restrict AI use. Each carries different implications for agency, fulfillment, and the central design challenge we identify across all three studies: ensuring that users retain a sense of authorship and meaningful contribution within AI-mediated work and life.

\section{Reflexivity and Limitations}

We approached this work as HCI researchers, not philosophers; our phenomenological approach is methodological rather than transcendental. Phenomenological methods produce richly textured accounts of lived experience, but they do not scale the way surveys or log analyses do. The toolkits we propose are designed for depth, not breadth; researchers seeking population-level trends will need to complement it with quantitative instruments. 

We see AI phenomenology not as replacing existing methods but as providing the experiential ground-truth that those methods often assume but rarely verify through triangulation of rich, qualitative information. Future work scaling these methods might derive closed-ended survey instruments from VAPT's value-chart comparison stages, develop automated coding schemes trained on our thematic analysis outputs, or embed experience-sampling probes within deployed AI systems to approximate longitudinal phenomenological data at scale---trading depth for breadth in ways that complement rather than supplant the qualitative core.

As AI systems become more capable, more autonomous, and more integrated into daily life, the question ``How did it feel?'' becomes as important as ``How well did it perform?'' The methods we develop now will shape what questions we can ask, and what experiences we can design, in the decades to come.

\bibliographystyle{ACM-Reference-Format}
\bibliography{bibliography}

\end{document}